\documentclass[prb,twocolumn,showpacs,preprintnumbers,amsmath,amssymb]{revtex4}
\usepackage[dvips]{graphicx}
\usepackage{graphicx}
\usepackage{dcolumn}
\usepackage{bm}
\usepackage{times}
\usepackage{mathptmx}
\usepackage{color}

\begin{document}

\title{Time-resolved imaging of magnetoelectric domain switching in multiferroic MnWO$_4$}

\author{T. Hoffmann,$^1$ P. Thielen,$^1$ P. Becker,$^2$ L. Bohat\'{y},$^2$ and M. Fiebig$^{1,3,\ast}$}

 \affiliation{$^1$HISKP, Universit\"{a}t Bonn, Nussallee 14-16, 53115 Bonn, Germany}
 \affiliation{$^2$Institut f\"{u}r Kristallographie, Universit\"{a}t zu K\"{o}ln, Z\"{u}lpicher Stra{\ss}e 49b, 50674 K\"{o}ln, Germany}
 \affiliation{$^3$Department of Materials, ETH Zurich, Wolfgang-Pauli-Strasse 10, 8093 Zurich, Switzerland}

\begin{abstract}
We show that the electric-field-induced reversal of the magnetic order parameter in multiferroic MnWO$_4$ occurs on the time scale of milliseconds. Throughout the switching process the magnetic order and the magnetically induced electric polarization remain rigidly coupled. The temporal progression of the domain structure was imaged with nanosecond
resolution by an electrical-pump--optical-probe technique using optical second harmonic generation. These nonequilibrium domain states significantly differ from the quasi-static domain reversal. A qualitative model gives an estimate of why the magnetoelectric order-parameter reversal in the
magnetically induced ferroelectrics is not inherently ultrafast.
\end{abstract}


\date{\today}

\maketitle

\section{Introduction: Time-resolved multiferroics}

Materials with cross-coupled magnetic and
electric properties, called magnetoelectrics, are intensely discussed because of their potential
for controlling dielectric or magnetic properties with the crosswise magnetic or electric field
.\cite{khomskii09} Magnetoelectric effects are intrinsically strong in compounds where magnetic long-range order breaks the inversion symmetry and induces a spontaneous polarization. This mechanism results in a rigid coupling of the improper ferroelectric polarization to the proper magnetic order parameter(s).\cite{newnham78,kimura03,hur04}

Despite intensive research on these so-called ``multiferroics'' one central aspect of the
interplay of magnetic and electric order parameters has scarcely been discussed: How does the
voltage-induced reversal of the magnetization progress with time? Knowing the dynamics of voltage-induced magnetization reversal would highlight the competition of
spins and charges. This would significantly advance our general understanding of the parameters defining the
magnetoelectric coupling in one of the most prominent classes of multiferroics. But also with
respect to possible applications like memory devices and sensors the switching dynamics is a key
issue. For example, the feasibility of any candidate compound for a novel type of memory element
depends crucially on its bit-reversal time.

Thus far, investigations of the dynamical properties of multiferroics are devoted to the AC
magnetoelectric response of magnetostrictive-piezoelectric composites at microwave frequencies.
\cite{fiebig05a} In two multiferroics the ultrafast response to laser pulse excitation has been
investigated.\cite{kimel01,matsubara09} Yet, in none of these experiments time-resolved
magnetoelectric switching processes were observed or discussed. In spite of the significance of
this topic investigations are virtually non-existent.

In this Report we show that the electric-field-induced reversal of the magnetic order in
multiferroic MnWO$_4$ extends over a surprisingly long time span in the order of milliseconds. The
evolution of the multiferroic domain structure throughout the switching process is spatially
resolved by optical second harmonic generation and compared to the quasi-static reversal of
domains. Fundamental differences between the observed magnetoelectric switching and models for
switching of domains in non-multiferroics are highlighted.

\section{Samples and methods}

\subsection{Multiferroic MnWO$_4$}

The compound investigated here, MnWO$_4$, is prototypical for the group of magnetically induced
ferroelectrics:\cite{arkenbout06,jia07} There is a multitude of similar compounds promoting
non-centrosymmetric, preferably frustrated and/or incommensurate spin structures. They can induce spontaneous polarization even up to ambient temperature.\cite{kimura08} This background emphasizes the general merit of our results.

Three magnetically ordered phases discussed in detail in Refs.~\onlinecite{lautenschlaeger93} and
\onlinecite{toledano09} constitute the phase diagram of MnWO$_4$. In the AF3 phase below
$T_N=13.5$~K the magnetic moments of Mn$^{2+}$ align collinearly along the easy axis while
their magnitudes are sinusoidally modulated yielding incommensurate order in the $xz$ plane (axes:
see Ref.~\onlinecite{axes}). An additional transverse spin component orders at $T_2=12.7$~K. In
this ${\rm AF3}\to{\rm AF2}$ transition the spin-density wave turns into an incommensurate
elliptical spin spiral that breaks the inversion symmetry and induces a spontaneous polarization
$P_y^{\rm sp}$ along the $y$ axis. Another transition at $T_1=7.6$~K leads to the AF1 phase with a
collinear centrosymmetric alignment of spins along the easy axis in the $xz$ plane. The
multiferroic AF2 phase is characterized by two magnetic domain states with opposite helicity of
the associated spin spiral. The helicity $\sigma_\pm$ is coupled to the direction of the
spontaneous polarization $\pm P_y^{\rm sp}$ so that both can be set or reversed by an electric
field $\pm E_y$.\cite{sagayama08}

\subsection{Optical second harmonic generation}

A convenient way for imaging the multiferroic domain structure is optical second harmonic
generation (SHG). Laser light is incident onto a crystal where it induces a frequency doubled wave
(Fig.~\ref{fig:static}(e)). The polarization of the SHG wave is determined by the symmetry of the
crystal and, hence, by any form of long-range order affecting the symmetry. Thus, SHG suggests
itself for probing \mbox{(anti-)} ferroic order of any type, especially in compounds with multiple
order parameters. In particular, the optical approach allows one to investigate the ordered state
with high temporal and spatial resolution \cite{fiebig05a,fiebig08} and resolve the
magnetoelectric switching of the domains. In previous SHG experiments, the as-grown multiferroic
domain structure in MnWO$_4$ was investigated \cite{meier09b} and the electric-field-induced
conversion into a single-domain state was observed in pyroelectric, neutron diffraction, and SHG
experiments.\cite{sagayama08,taniguchi06,meier09a} However, in none of these experiments the
actual poling process with its temporally evolving domain structure was resolved.

\subsection{Experimental details} \label{sec:exdetail}

The electric field was applied along the $y$ axis of the samples. As shown in the insets of
Fig.~\ref{fig:dynamic}, this is achieved for the (010) samples by glass plates coated with 100~nm
of transparent indium-tin-oxide, whereas for the (100) samples polished steel electrodes are used.
The quasi-static evolution of the domain structure in Fig.~\ref{fig:static} was investigated by
applying the electric field to the electrodes with a tunable high-voltage supply. The dynamic
reversal of the domains in Figs.~\ref{fig:dynamic} to \ref{fig:fastLocal} was resolved with the
electrical-pump--optical-probe technique depicted in Fig.~\ref{fig:dynamic}(a). Repetitive
rectangular voltage pulses are applied to the samples. At time zero the voltage pulse reverses the
electric field applied to the sample from $-E_y$ to $+E_y$ as indicated by the bar on top of
Figs.~\ref{fig:dynamic}(a), \ref{fig:dynamic}(b), and \ref{fig:dynamic}(c). The slope of the
voltage pulse is 50~ns. Different fields up to $E_y=\pm900$~kV/m were used. The state of the sample is probed by SHG at the time $\Delta t$ while the electric
field pulse is still applied. After 50~ms the electric field is reversed to $-E_y$ again. However, this exceeds the time
span displayed in our figures so that we only see the initial reversal of the electric field at $t=0$. SHG data are accumulated across about 1000 laser pulses and, thus, voltage
cycles per data point.

The samples are placed in an optical cryostat. The temporal evolution of the magnetic system is
tracked by changing the delay $\Delta t$ between the voltage reversal at $t=0$ and the 5-ns laser
pulses (energy: $1-5$~mJ) as source of the SHG. The transmission setup for magnetically induced
SHG and a comprehensive characterization of SHG on multiferroic MnWO$_4$ are detailed in
Refs.~\onlinecite{fiebig05a} and \onlinecite{meier10}. Summarizing these, the SHG experiments were done on
MnWO$_4$(010) samples in which the $\chi^{\rm ED}_{x^{\prime}z^{\prime}z^{\prime}}$ contribution
at 2.21~eV couples to the incommensurate magnetic order but not to the spontaneous polarization.
\cite{axes,meier10} In addition, the magnetically induced polarization was probed via the $\chi^{\rm ED}_{yzz}$ contribution at 2.72~eV in MnWO$_4$(100)
samples.\cite{meier10} All samples were grown by top-seeded melt growth and lapped and
etch-polished to a thickness of $0.4-1.0$~mm with a silica slurry. A summary of the samples, SHG
contributions, and types of order probed in our experiments is given in Table~\ref{tab:SHGsummary}.

\section{Results and discussion}

\subsection{Quasi-static switching of domains}

\begin{figure}[h]
\centering
\includegraphics[width=8.7cm,keepaspectratio,clip]{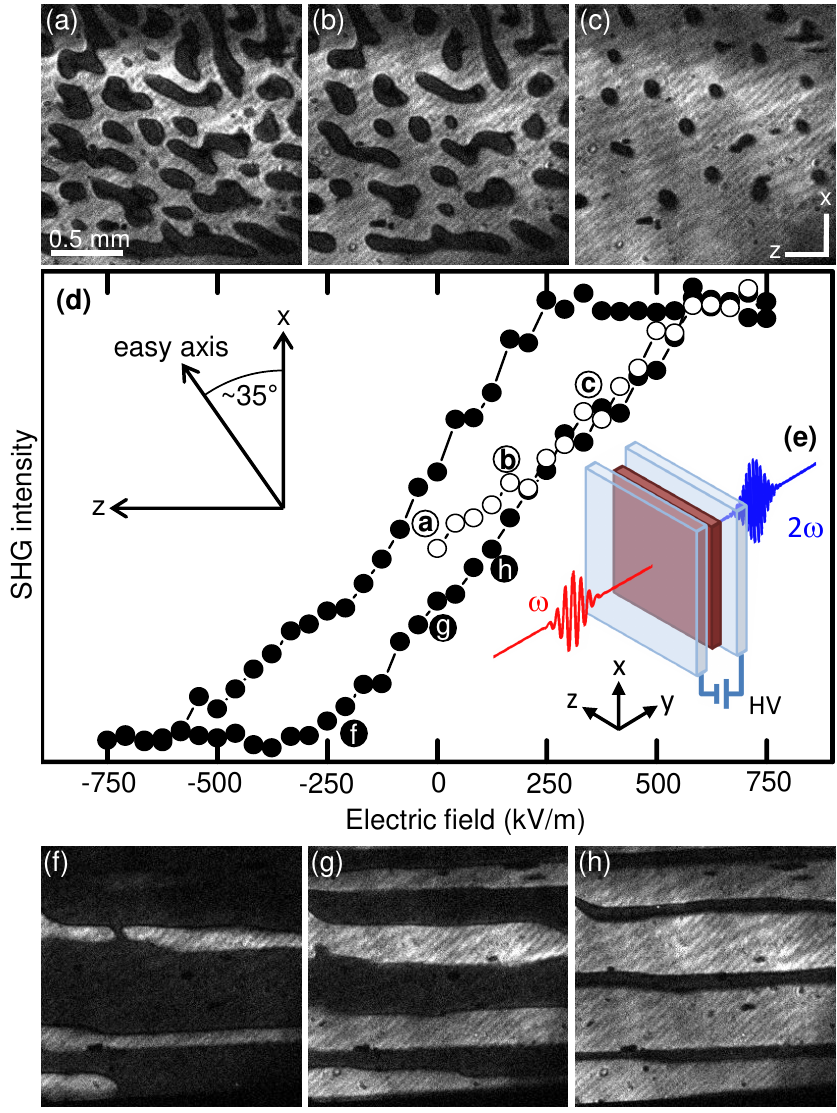}
\caption{\label{fig:static}(Color online) Evolution of the magnetic domain structure in multiferroic
MnWO$_4$(010) during quasi-static electric-field poling. (a--c) Poling from a multidomain state
after zero-field cooling. (d) SHG hysteresis loop with open symbols denoting the virgin curve. (e)
Sample environment with applied voltage (HV) and incident fundamental and emitted SHG light.
(f--h) Poling from a single-domain state at $-750$~kV/m during gradual increase of the field
towards $+750$~kV/m. Bright and dark areas in the SHG images correspond to domains with opposite
magnetic helicity. Temperature is 11~K.}
\end{figure}

Figure~\ref{fig:static} shows the quasi-static switching of magnetic helicity domains in
MnWO$_4$(010) by an electric field. Starting from a zero-field-cooled state ($E_y=0$) or a
single-domain state ($E_y=-750$~kV/m), the electric field in Figs.~\ref{fig:static}(a) to
\ref{fig:static}(c) and Figs.~\ref{fig:static}(f) to \ref{fig:static}(h), respectively, is
increased to $E_y=+750$~kV/m with a time $\geq 1$~s between subsequent data points. The
corresponding virgin curve in Fig.~\ref{fig:static}(d) is followed by a hysteresis loop that is
well centered in contrast to the loops in Ref.~\onlinecite{finger10}. The coercive field is
120~kV/m and the largest field applied exceeds the saturation field of $\pm 600$~kV/m by 25\%. The
domain structure obtained after zero-field cooling was already discussed in
Ref.~\onlinecite{meier09b}: A magnetic bubble topology emerges. On the surface of the (010) sample
the domain boundaries yield an affinity to be oriented along the $z$ axis and the magnetic easy
axis. Domain states of opposite helicity differ in their brightness because of SHG interference
effects.\cite{meier09a} In the present work we see that in the increasing electric field the
bubbles for either of the domain states shrink with a tendency to minimize the surface. This leads
to the spherical domain shape visible on the (010) face in Fig.~\ref{fig:static}(c). Finally, a
single-domain state is acquired.

In stark contrast to the bubble topology the re-emergence of the opposite domain state during
gradual reversal of the poling field leads to the formation of domains that appear as stripes
along the $z$ axis on the surface of the (010) sample, see Figs.~\ref{fig:static}(f) to
\ref{fig:static}(h). Nucleation of domains occurs primarily at the sample edges but was also
observed within the sample. Note that the direction of the stripes is not that of the spontaneous
polarization or of the electric poling field.

As an interpretation, the nucleation of domains in Figs.~\ref{fig:static}(a) to
\ref{fig:static}(c) is determined by the emergence of a spin spiral upon crossing $T_2$. This is
an entirely magnetic phenomenon because the primary order parameter is magnetic and electric fields are not involved. In contrast, the domain
structure in Figs.~\ref{fig:static}(f) to \ref{fig:static}(h) results from the modification of a
spin spiral that is already present. This modification is based on magnetoelectric interactions
because it is driven by the electric field. Note that the weakest magnetic exchange is found along
the $x$ direction.\cite{tian09} This explains the formation of the domain stripes along z on the
(010) face.

\subsection{Dynamic switching of domains}

\begin{figure}[h]
\centering
\includegraphics[width=8.7cm,keepaspectratio,clip]{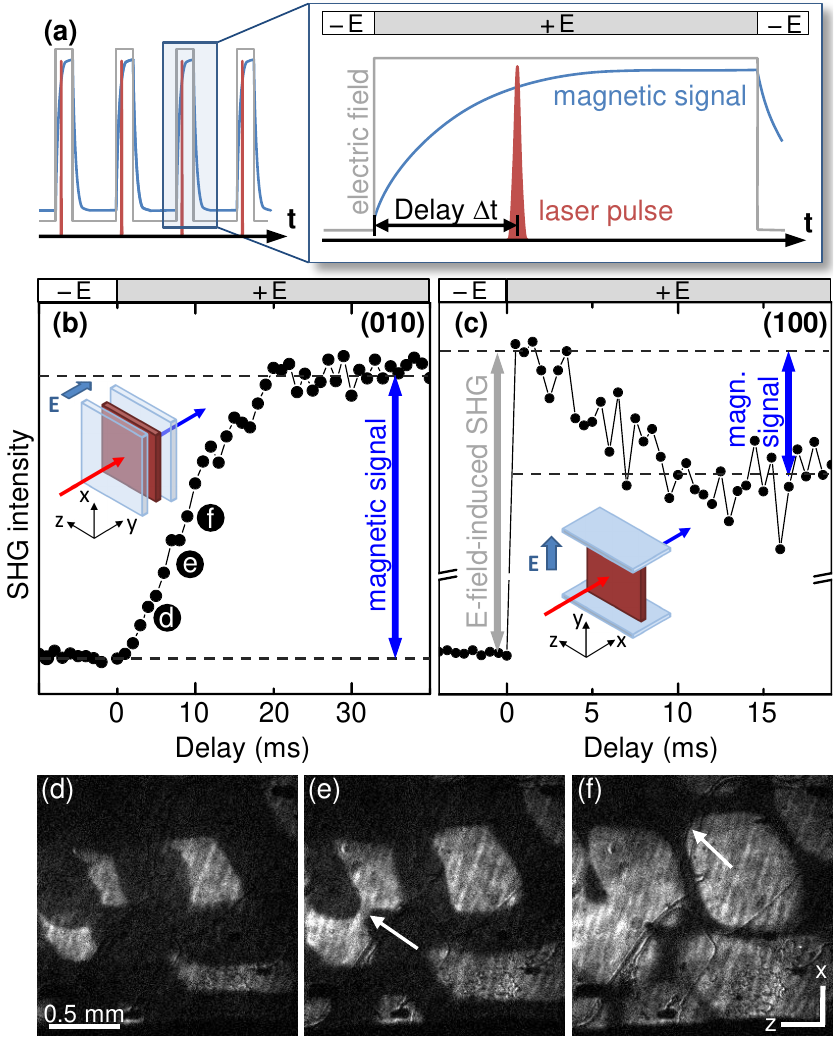}
\caption{\label{fig:dynamic}(Color online) Evolution of the domain structure during dynamic
electric-field poling with a voltage pulse ($E=750$~kV/m, slope 50~ns, width 50~ms). (a) Sketch
of the repetitive electric-field poling with subsequent SHG probing. (b, c) SHG intensity as a
function of the delay to the voltage pulse at $t=0$ for (b) MnWO$_4$(010) 11.8~K, 750~kV/m and (c)
MnWO$_4$(100) 12.0~K, 330~kV/m. Insets show the position of the electrodes with respect to the
laser beam. The dashed lines show the SHG intensity for the two single domain states. Note that the abrupt jump in (c) at $t=0$ is caused by EFISH and is thus not related to the long range order. (d--f) Magnetic domain structure at various
delays. Arrows: see text. Bars on top of (a--c) indicate the applied electric field.}
\end{figure}

Figure~\ref{fig:dynamic} shows the dynamic switching of magnetic helicity domains in MnWO$_4$(010)
by an electric field. As mentioned in Section~\ref{sec:exdetail} the fastest time in which the electric field can be reversed is 50~ns. This is determined by the slope of the
repetitively applied voltage pulses. Figure~\ref{fig:dynamic}(b) reveals that the response of the
magnetic order parameter to the electric field pulses of $E_y=750$~kV/m is strikingly slow: The
reversal of the magnetic domain state takes as long as 20~ms. The domain structure throughout this
dynamic reversal is shown in Figs.~\ref{fig:dynamic}(d) to \ref{fig:dynamic}(f). It differs substantially from the one in zero-field-cooled samples and also from the
one obtained during quasi-static field reversal (Fig.~\ref{fig:static}). The dynamic reversal leads to the nucleation of
rhomboid-shaped domains whose sides run approximately parallel to the $z$ axis and the magnetic
easy axis. In this nonequilibrium state, a tendency to minimize the domain surface is already
present in the form of rounded corners (arrow in Fig.~\ref{fig:dynamic}(f)) and domain fusion
(arrow in Fig.~\ref{fig:dynamic}(e)) but much less pronounced than in Fig.~\ref{fig:static}(c).
Although the images are accumulated from about 1000 laser pulses (i.e.\ 1000 poling cycles) a
well-defined domain structure is obtained. Therefore, the domain reversal always progresses
through the same domain pattern. This pronounced memory effect is a strong indication for pinning
effects.

The same type of dynamic switching experiment was also carried out on a
MnWO$_4$(100) sample with the result shown in Fig.~\ref{fig:dynamic}(c). In this geometry the
electric field is limited to a maximum value of $\pm 330$~kV/m because of the larger distance
between the electrodes and the limited voltage of the HV source.
At $t<0$ the SHG intensity remains constant for an electric field of $-330$~kV/m. The electric-field
reversal to $330$~kV/m at $t=0$ leads to a \textit{jump} of the SHG signal within the rise time of the voltage pulse.
This is followed by a gradual \textit{decrease} of the SHG intensity lasting until $t=10$~ms. At
$t>10$~ms, the SHG intensity remains constant at a level that is different from that at $t<0$.
Upon the field reversal from $+330$ to $-330$~kV/m (not shown) the SHG intensity returns to its
original value.

The jump at $t=0$ is caused by SHG contributions that are induced by the externally applied field
but unrelated to the long-range order.\cite{meier09a} This effect is well-known and documented as
``EFISH'' (electric-field-induced second harmonic) in the literature.\cite{he00} The EFISH signal
follows the applied voltage without delay so that its rise time is 50~ns in
Fig.~\ref{fig:dynamic}(c). EFISH contributions are always polarized parallel to the applied
electric field which explains their absence in Fig.~\ref{fig:dynamic}(b) where the light is
propagating along the field direction. Only the decrease of the SHG signal from the upper to the lower
dashed line in Fig.~\ref{fig:dynamic}(c) is associated to the actual domain reversal. The reversal
changes the domain-related SHG contribution. This changes the interference with the (constant)
EFISH contribution and, hence, the net SHG yield until it stabilizes after the domain reversal is
completed at $t=10$~ms. The factor of two in switching time between Figs.~\ref{fig:dynamic}(b) and
\ref{fig:dynamic}(c) originates from differences in the sample temperature and the applied
electric fields, as discussed in detail below.

As mentioned, the (100) sample promotes SHG contributions coupling to the spontaneous polarization.\cite{meier10} Any transient decoupling of the polarization from the magnetization would
therefore manifest in Fig.~\ref{fig:dynamic}(c) as a timescale different from that of the
magnetization in Fig.~\ref{fig:dynamic}(b). In particular, a fast response was considered likely
for any electronic contributions to the spontaneous polarization.\cite{picozzi07,lottermoser09}
However, no such decoupling is observed. (Remember that the time scale introduced by the jump at
$t=0$ is related to the applied electric field and \textit{not} to a switching of the multiferroic
order.) We thus conclude that the magnetic and electric order parameter remain rigidly clamped
throughout the switching process so that the polarization, just like the magnetization, reverses
on a millisecond time scale. Since faster switching processes are not present it is not
necessary to increase the time resolution of the experiment beyond the nanosecond range employed
here.

\begin{figure}[h]
\centering
\includegraphics[width=8.7cm,keepaspectratio,clip]{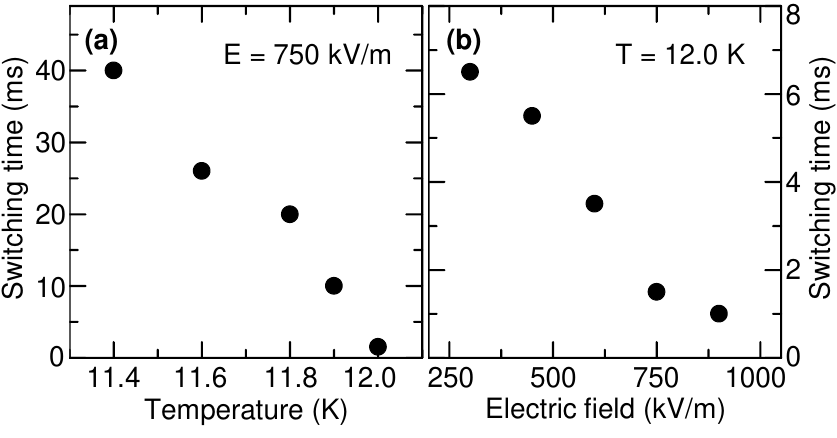}
\caption{\label{fig:switchingtime}Dependence of the magnetoelectric domain-reversal time on (a)
temperature and (b) electric field.}
\end{figure}

In order to investigate variations of the magnetoelectric domain-reversal time $\tau$ dynamic
switching experiments as in Fig.~\ref{fig:dynamic}(b) we carried out at different temperatures and
electric fields. The accessible temperature interval is limited to 0.6~K by the increase of the coercive
field towards lower and the decrease of the SHG intensity towards higher temperature.
The electric field is determined by the field required to acquire a single-domain state and is limited to 900~kV/m by the breakthrough field in helium gas. Figure~\ref{fig:switchingtime} shows the reversal time in dependence of these parameters, probed
via the magnetic SHG signal on the (010) sample. First, Fig.~\ref{fig:switchingtime}(a) reveals
that $\tau$ increases with decreasing temperature. Although a temperature interval of only 0.6~K 
is investigated $\tau$ changes by a factor of 40, from 1~ms at 12.0~K to 40~ms at
11.4~K in an electric field of 750~kV/m. Figure~\ref{fig:switchingtime}(b) shows a series of
experiments at 12.0~K with fields between 300~kV/m (the saturation field at 12.0~K) and
900~kV/m. Clearly, $\tau$ decreases with increasing field. However, due to the limited field
range we cannot make a conclusive statement on the validity of Merz's law \cite{merz57}, which describes the field dependence of the switching time in ferroelectrics. It predicts a relation
$\tau \propto\exp(-E/E_a)$ with $E_a$ as activation field.

\begin{figure}[h]
\centering
\includegraphics[width=8.7cm,keepaspectratio,clip]{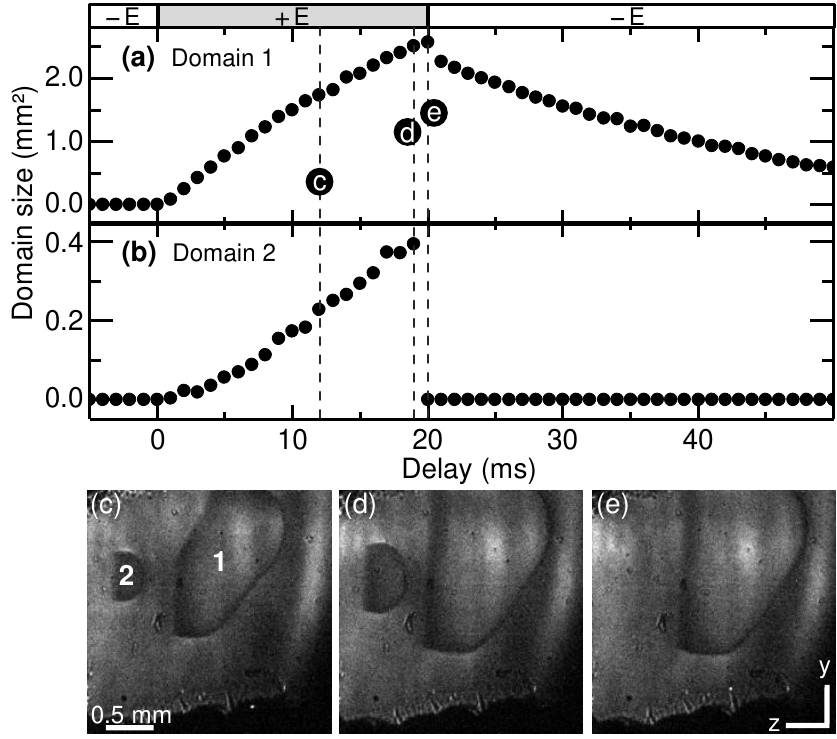}
\caption{\label{fig:fastLocal} Evolution of the domain structure in MnWO$_4$(100) during repeated
electric-field poling with $E=\pm 330$~kV/m at 11.5~K. (a, b) Area of domains 1 and 2 as a function
of the delay to the first voltage pulse. The bar on top indicates the applied electric field.
(c--e) SHG images of the domains at the given delays.}
\end{figure}

In Fig.~\ref{fig:fastLocal} the evolution of two selected domains in a MnWO$_4$(100) sample for a
scenario with two field reversals at $t=0$ and at $t=20$~ms is investigated. The area of the
domains is shown in dependence of the delay after the first field reversal; the respective
electric field is indicated by the bar on top of Fig.~\ref{fig:fastLocal}(a). The timing of the
voltage pulses was chosen such that the second field reversal occurs before the first domain
reversal has been completed. After the initial voltage reversal at $t=0$ the areas of domains 1
and 2 grow steadily, albeit at different rates. After the second voltage reversal, domain 1
shrinks again, with a small discontinuity right at the voltage reversal. In contrast, domain 2
disappears abruptly, and unrelated to the voltage reversal. Hence, although the magnetoelectric
domain switching progresses smoothly according to Fig.~\ref{fig:dynamic}, Fig.~\ref{fig:fastLocal}
reveals that local deviations from the averaged response are possible. However, these local
deviations are not predictable or controllable and they change with each annealing cycle applied
to the sample. Note that the collapse of domain 2 shows that the \textit{intrinsic} speed of
domain switching is orders of magnitude faster than the \textit{average} response of the domains
to the electric field. Collapsing domains were observed on the (010) as well as on the (100) face
of the sample with collapse times faster than 25~ns, the smallest delay between two images
investigated in this experiment.

\section{Discussion of the magnetoelectric switching}

Summarizing Figs.~\ref{fig:static} to \ref{fig:fastLocal} we conclude that the reversal of
magnetic domains by an electric field is governed by domain wall movements on a timescale in the
order of milliseconds and thus a surprisingly slow process. Pronounced memory effects govern the
reversal. The topology of the magnetic helicity domains was investigated under three different
conditions: (i) emergence of the helix (zero-field cooling through $T_2$), and reversal of the
existing helix under (ii) equilibrium conditions (quasi-static electric-field-induced reversal)
and (iii) non-equilibrium conditions (dynamic reversal).

We now have to compare the switching here observed in a multiferroic with that of ferroelectric or
magnetic domain reversal by the adjunct electric or magnetic field, respectively, in
non-multiferroics. This will highlight some of the principal differences between the
magnetoelectric switching processes in this work and the well-documented switching of, e.g.,
ferroelectric domains by electric fields. There is no explicit theory on time-resolved
magnetoelectric domain reversal so that we take a first qualitative approach here.

Models for the reversal of ferroelectric domains involve nucleation at defects, fast growth along
the field direction, and slow growth perpendicular to the electric field. \cite{scott00} The
``fast forward growth'' occurs at the speed of sound and is related to the local charge at the
domain walls perpendicular to the spontaneous polarization. In Fig.~\ref{fig:fastLocal} the
forward growth is absent. Domain walls parallel and perpendicular to the poling field propagate at
about 1~m/s and, hence, at a speed three orders of magnitude below that of sound. We suggest that
this is related to the smallness of the spontaneous polarization in the magnetically-induced
ferroelectrics \cite{khomskii09,toledano09} and the resulting dominance of other mechanisms
determining domain-wall propagation. One such mechanism is the coupling of the spontaneous
polarization to the spin system, and Fig.~\ref{fig:dynamic} revealed that this coupling is rigid.
Therefore, the energies competing in the magnetoelectric domain reversal are the electric-field
energy $\cal{E}$ and the magnetic anisotropy energy $K$. Here an orders-of-magnitudes estimate
reveals $K/{\cal{E}}\approx 100$ which corroborates the inertness of the magnetic system for the
electric field acting on it: The electric-field energy is orders of magnitude weaker than the
energy barrier separating the states with opposite magnetic helicity.

The resulting
magnetoelectric nature of the switching process is reflected by the manifestation of the magnetic
easy direction in the dynamic domain structure in Figs.~\ref{fig:dynamic}(d)-(f) which is much
more pronounced than in Fig.~\ref{fig:static}. Measurement of the electric-field dependence of the
domain reversal beyond the saturation field, including a verification of the validity of Merz's
law,\cite{merz57} could yield further insight but this is restricted by the low breakthrough
voltage of the helium gas cooling the sample.

\section{Summary and conclusion}

In summary, time-resolved SHG experiments on multiferroic MnWO$_4$ crystals revealed that the
electric-field-induced reversal of the magnetization in multiferroics with magnetically induced
improper ferroelectricity is inherently {\it not} ultrafast. On the one hand, rapid reversal is
inhibited by the inertness of the magnetic system in the electric field which is expressed by the
smallness of the electric-field energy $\cal{E}$ in comparison to the magnetic anisotropy energy.
On the other hand, the expectation that fast electronic contributions to the polarization
\cite{picozzi07,lottermoser09} might transiently decouple from the slowly reversing magnetization
was not confirmed: According to our experiments the ferroelectric polarization remains rigidly
clamped to the magnetization throughout the switching process.

With respect to applications, for magnetoelectrically controlled giant-magnetoresistance sensors
\cite{ramesh07} the electric-field induced reversal of the magnetic order parameter does not have
to be faster than observed here. For these applications Figs.~\ref{fig:static} to \ref{fig:fastLocal} show that
memory effects result in a highly reproducible magnetoelectric switching process. No signs of
fatigue were observed in our experiments. High durability may therefore be a major advantage of
magnetoelectric devices. In addition, understanding and control of the pinning mechanisms causing
the memory effect will be the key for tailoring the magnetically induced ferroelectrics according
to technological requirements.

Finally, there is a clear need for a theory modelling the non-equilibrium dynamics of the
voltage-induced reversal of magnetic domains in multiferroics like MnWO$_4$. Here our
orders-of-magnitude estimate of the energy scales involved in the switching process provides a
clue about the approach such a theory might take.

The authors thank N. A. Spaldin for helpful discussions and the DFG (SFB 608) for financial
support.

\begin{table*}
\begin{tabular*}{\textwidth}{@{\extracolsep{\fill}} llll}
Sample orientation & SHG component & Order probed & EFISH \\
\hline
(010) & $\chi^{\rm ED}_{x^{\prime}z^{\prime}z^{\prime}}$ @ 2.21~eV & magnetic & no \\
(100) & $\chi^{\rm ED}_{yzz}$ @ 2.72~eV & magnetic \& electric order & yes \\
\hline
\end{tabular*}
\caption{\label{tab:SHGsummary}Summary of the experimental configurations used in our experiments.
Columns correspond to: the orientation of the sample, the SHG contributions probed (tensor
component and photon energy), the type of order probed, the presence of EFISH contributions.}
\end{table*}





\begin{thebibliography}{99}

\bibitem[*]{dummy} Email address: manfred.fiebig@mat.ethz.ch

\bibitem{khomskii09}
D. Khomskii,
Physics {\bf 2}, 20 (2009).

\bibitem{newnham78}
R. E. Newnham J. J. Kramer, W. A. Schulze, and L. E. Cross,
J.\ Appl.\ Phys.\ {\bf 49}, 6088 (1978).

\bibitem{kimura03}
T. Kimura, T. Goto, H. Shintani, K. Ishizaka, T. Arima, and Y. Tokura,
Nature {\bf 426}, 55 (2003).

\bibitem{hur04}
N. Hur, S. Park, P. A. Sharma, J. S. Ahn, S. Guha, and S. W. Cheong,
Nature {\bf 429}, 392 (2004).

\bibitem{fiebig05a}
M. Fiebig, V. V. Pavlov, and R. V. Pisarev,
J.\ Opt.\ Soc.\ Am.\ B {\bf 22}, 96 (2005).

\bibitem{kimel01}
A. V. Kimel, R. V. Pisarev, F. Bentivegna, and Th.\ Rasing,
Phys.\ Rev.\ B {\bf 64}, 201103(R) (2001)

\bibitem{matsubara09}
M. Matsubara, Y. Kaneko, J.-P. He, H. Okamoto, and Y. Tokura,
Phys.\ Rev.\ B {\bf 79}, 140411(R) (2009).

\bibitem{arkenbout06}
A. H. Arkenbout, T. T. M. Palstra, T. Siegrist, and T. Kimura,
Phys.\ Rev.\ B {\bf 74}, 184431 (2006).

\bibitem{jia07}
C. Jia, S. Onoda, N. Nagaosa, and J. H. Han,
Phys.\ Rev.\ B {\bf 76}, 144424 (2007).

\bibitem{kimura08}
T. Kimura, Y. Sekio, H. Nakamura, T. Siegrist and A. P. Ramirez,
Nat.\ Mater.\ {\bf 7}, 291 (2008).

\bibitem{lautenschlaeger93}
G. Lautenschl\"{a}ger, H. Weitzel, T. Vogt, R. Hock, A. B\"{o}hm, M. Bonnet, and H. Fuess,
Phys.\ Rev.\ B {\bf 48}, 6087 (1993).

\bibitem{toledano09}
P. Tol\'{e}dano, B. Mettout, W. Schranz, G. Krexner,
J. Phys.: Condens. Matter {\bf 22} 065901 (2010).

\bibitem{axes}
The Cartesian system ($\mathbf x$, $\mathbf y$, $\mathbf z$) is related to the monoclinic
crystallographic system ($a$, $b$, $c$ with $a=4.830$~\AA, $b=5.7603$~\AA, $c=4.994$~\AA, and
$\beta=91.14^{\circ}$) as follows: $\mathbf{x}=\mathbf{a}/a$, $\mathbf{y}=\mathbf{b}/b$,
$\mathbf{z}=\mathbf{x}\times\mathbf{y}$. Note that because of $\beta\approx 90^{\circ}$
$\mathbf{z}$ is pointing approximately along $\mathbf{c}$. The rotated Cartesian system (${\mathbf
x}^{\prime}$, ${\mathbf y}^{\prime}$, ${\mathbf z}^{\prime}$) is defined by the incommensurate
structure as detailed in Ref.~\protect\onlinecite{meier10}.

\bibitem{sagayama08}
H. Sagayama, K. Taniguchi, N. Abe, T. H. Arima, M. Soda, M. Matsuura, and K. Hirota,
Phys.\ Rev.\ B {\bf 77}, 220407(R) (2008).

\bibitem{fiebig08}
M. Fiebig {\it et al.}, 
M. Fiebig, N. P. Duong, T. Satoh, B. B. Van Aken, K. Miyano, Y. Tomioka, and Y. Tokura,
J.\ Phys.\ D {\bf 41}, 164005 (2008).

\bibitem{meier09b}
D. Meier, N. Leo, M. Maringer, T. Lottermoser, M. Fiebig, P. Becker, and L. Bohat\'{y},
Phys.\ Rev.\ B {\bf 80}, 224420 (2009).

\bibitem{taniguchi06}
K. Taniguchi, N. Abe, T. Takenobu, Y. Iwasa, and T. Arima,
Phys.\ Rev.\ Lett.\ {\bf 97}, 097203 (2006).

\bibitem{meier09a}
D. Meier, M. Maringer, Th.\ Lottermoser, P. Becker, L. Bohat\'{y}, and M. Fiebig,
Phys.\ Rev.\ Lett.\ {\bf 102}, 107202 (2009).

\bibitem{meier10}
D. Meier, N. Leo, G. Yuan, Th.\ Lottermoser, M. Fiebig, P. Becker and L. Bohat\'{y}
Phys.\ Rev.\ B {\bf 82}, 155112 (2010).

\bibitem{finger10}
T. Finger, D. Senff, K. Schmalzl, W. Schmidt, L. P. Regnault, P. Becker, L. Bohat\'{y} and M. Braden,
Phys.\ Rev.\ B {\bf 81}, 054430 (2010).

\bibitem{tian09}
C. Tian, C. Lee, H. Xiang, Y. Zhang, C. Payen, S. Jobic, and M.-H. Whangbo,
Phys. Rev. B 80, 104426 (2009).

\bibitem{he00}
G. S. He and S. H. Liu, {\it Physics of Nonlinear Optics}, (World Scientific, Singapore, 2000).

\bibitem{picozzi07}
S. Picozzi, K. Yamauchi, B. Sanyal, I.A. Sergienko, and E. Dagotto
Phys.\ Rev.\ Lett.\ {\bf 99}, 227201 (2007).

\bibitem{lottermoser09}
Th.\ Lottermoser, D. Meier, R. V. Pisarev, and M. Fiebig,
Phys. Rev. B {\bf 80}, 100101(R) (2009).

\bibitem{merz57}
W. J. Merz,
Phys.\ Rev.\ {\bf 95}, 690 (1954).

\bibitem{scott00}
J. F. Scott, {\it Ferroelectric Memories}, (Springer, Berlin, 2000).

\bibitem{ramesh07}
R. Ramesh and N. A. Spaldin,
N.\ Mater.\ {\bf 6}, 21 (2007).

\end{thebibliography}
\end{document}